\documentclass[a4paper, 11pt]{article}
\usepackage[english]{babel}
\makeatletter

\usepackage{geometry,graphicx,color}
\usepackage{amsfonts, amsmath, amssymb, slashed,dsfont}
\usepackage{tensor}
\usepackage{hyperref}
\usepackage{epsfig}
\usepackage{pifont}
\usepackage{enumitem}
\usepackage{csquotes}
\usepackage{bm}
\usepackage{mathrsfs} % Allows \mathscr
\usepackage[hyperref,thmmarks,amsmath]{ntheorem}
\usepackage{amsmath,amssymb,slashed}
\usepackage{comment}
\usepackage{tikz}
\usetikzlibrary{positioning}
\usetikzlibrary{decorations.markings}
\usetikzlibrary{arrows.meta}
\usepackage{filecontents}

\usepackage[natbib=true,sorting=none]{biblatex}
\numberwithin{equation}{section}
\newcommand\bbone{\mathbb{I}}

\newcommand\kS{{\mathfrak S}}

\newcommand{\sign}{\mathrm{sign}}
\newcommand\dd{\mathrm{d}}

\newcommand\sstar{\text{\ding{86}}}
\newcommand{\institute}[1]{\newcommand{\@institute}{#1}}
\renewcommand{\maketitle}{
\vspace*{0.5\baselineskip}
{% title
\center\LARGE\noindent\@title\par
}%
\vspace{1.5\baselineskip}
{% author
\center\normalsize\noindent\ignorespaces\@author\par
}%
\vspace{0.5\baselineskip}
{% institutequantum
\center\normalsize\ignorespaces\@institute\par
}%
\vspace{2\baselineskip}
}%

\begin{document}

% -------------- Title -----------------
\title{Gauge theory on $\rho$-Minkowski space-time}
\author{Valentine Maris$^{a,b}$, Jean-Christophe Wallet$^a$}
\institute{%
\textit{$^a$IJCLab, Universit\'e Paris-Saclay, CNRS/IN2P3, 91405 Orsay, France}\\
\textit{$^b$Univ Lyon, ENS de Lyon, Univ Claude Bernard, CNRS, Laboratoire de Physique, 69342 Lyon, France}
\bigskip\\

e-mail:  
\href{mailto:valentine.maris@ijclab.in2p3.fr}{\texttt{valentine.maris@ijclab.in2p3.fr}}

\href{mailto:jean-christophe.wallet@universite-paris-saclay.fr}{\texttt{jean-christophe.wallet@universite-paris-saclay.fr}}
}%
\maketitle

%---------------------- Abstract --------------------
\begin{abstract} 
We construct a gauge theory model on the 4-dimensional $\rho$-Minkowski space-time, a particular deformation of the Minkowski space-time recently considered. The corresponding star product results from a combination of Weyl quantization map and properties of the convolution algebra of the special Euclidean group. We use noncommutative differential calculi based on twisted derivations together with a twisted notion of noncommutative connection. The twisted derivations pertain to the Hopf algebra of $\rho$-deformed translations, a Hopf subalgebra of 
the $\rho$-deformed Poincar\'e algebra which can be viewed as defining the quantum symmetries of the $\rho$-Minkowski space-time. The gauge theory model is left invariant under the action of the $\rho$-deformed Poincar\'e algebra. The kinetic part of the action is found to coincide with the one of the usual (commutative) electrodynamics.

\end{abstract}

%\tableofcontents
\newpage
\section{Introduction}\label{sec:intro}
Quantum gravity aims to provide a suitable characterisation of the gravitational interaction at very short distances and very high energy. Recent studies on extreme cosmological events and exploring the quantum regime may provide a way to set-up possible observational tests. See for instance \cite{zerevue}, \cite{zewhitepaper}. 
This challenging theoretical question has been approached from different viewpoints giving rise to a large amount of works, from which a consensus often shows up that Quantum Gravity may well give rise to a quantum space-time in some 
effective regime.\\ 

It appears that quantum space-times, for which most of the usual notions linked to manifolds no longer make sense, can be conveniently described within the framework of noncommutative geometry \cite{connes}. Many examples of these quantum objects are now available in the physics literature, among which some physically promising ones are those acted on by a deformation of the Poincar\'e symmetry, interpreted as the "quantum space-time symmetry" with deformation parameter identified with the Planck mass or eventually the scale of Quantum Gravity. In the vein of the development of field theories on these quantum spaces called generically noncommutative field theories, the construction of related gauge model versions has been the subject of an intense activity. For a review, see \cite{physrep}. \\

Gauge theory models on the Moyal spaces, a noncommutative structure possibly emerging in String Theory \cite{Seiberg_1999}, \cite{Schomerus_1999},  have been considered a long ago either as noncommutative extension of Yang-Mills type theories or as matrix models, see e.g. in \cite{Mart_n_2001}\nocite{Blaschke_2010}\nocite{Blaschke_2013}\nocite{de_Goursac_2007}\nocite{de_Goursac_2008}\nocite{Aoki_2000}\nocite{Steinacker_2007}\nocite{Madore_2000}-\cite{Grosse_2008}. Gauge theories on $\mathbb{R}^3_\lambda$, a deformation of the 3-dimensional Euclidean space{\footnote{For earlier work on $\mathbb{R}^3_\lambda$, see e.g. \cite{Hammou_2002}, \cite{Gracia_Bond_a_2002}.}} which shows up for instance in Group Field Theory, see e.g. \cite{oriti2009group}, have also been considered mainly from the viewpoint of matrix models \cite{Steinacker_2004}\nocite{Castro_Villarreal_2005}\nocite{G_r__2014}\nocite{G_r__2015}-\cite{Wallet_2016}. The $\kappa$-Minkowski space-time, rigidely linked through a duality to
a deformation of the Poincar\'e algebra called the $\kappa$-Poincar\'e algebra, appeared three decades ago \cite{luk-ruegg}, \cite{majid1} and has now acquired a physically prominent place among the quantum spaces \cite{Lukierski_2017}, for instance as providing a realisation of the Double Special Relativity \cite{Amelino_Camelia_2002}, \cite{Kowalski_Glikman} or for its possible relationship to Relative Locality \cite{Amelino_Camelia_2011}, \cite{Gubitosi_2013}. Field theories and gauge theories on the $\kappa$-Minkowski space-time have been considered in \cite{marija11}\nocite{marija13}\nocite{gross-wohl}\nocite{marija12}\nocite{PW2018}\nocite{PW2019}\nocite{MW2020a}-\cite{MW2021}.\\

In this paper, we will focus on another deformation of the Minkowski space-time, called the $\rho$-Minkowski space-time, first considered almost two decades ago in \cite{Lukierski_2006}. Similarly to the 
$\kappa$-Minkowski space-time, this recently (re)considered quantum space-time is acted on by a deformation of the Poincar\'e algebra, hence called the $\rho$-Poincar\'e algebra. The relevance of this so far poorly explored quantum space-time has been examined in black-hole physics in \cite{marija1}, \cite{marija2} and its impact on localisability and quantum observers studied in \cite{localiz1}, \cite{localiz2}. For appearance in ADS/CFT context, see \cite{Meier:2023kzt}, \cite{Meier:2023lku} . The algebraic structures underlying the $\rho$-deformed Poincar\'e symmetry have been explored recently in \cite{fabiano2023bicrossproduct}, showing in particular the isomorphism between the bicrossproduct structure and the Drinfeld twist approach of the deformed symmetry, based on the Drinfeld twist first defined in \cite{Lukierski_2006}. Some quantum one-loop properties of ($\rho$-Poincar\'e invariant) scalar field theories with quartic interaction based on different star-products were studied in \cite{Dimitrijevi_iri__2018} and \cite{rho-1}. The overall conclusion of both works is that UV/IR mixing occurs in these field theories.\\

The purpose of this paper is to build the action for a gauge theory on the (four-dimensional) $\rho$-Minkowski space-time, using for that purpose a noncommutative differential calculus based on twisted derivations. The star-product modeling 
the $\rho$-Minkowski space-time is the one used in \cite{rho-1}. The invariance of the classical action under the $\rho$-deformation of the Poincar\'e symmetry is examined.\\

The paper is organised as follows. In Section \ref{sec:section2}, the twisted noncommutative differential calculus is characterized. In Section \ref{section3}, we define the twisted (hermitian) connection and related curvature together with the gauge transformations. In Section \ref{section4}, a gauge invariant action whose commutative limit coincides with the usual action for electrodynamics is presented and discussed. The $\rho$-deformation of the Poincar\'e (Hopf) algebra leaving this action invariant is characterized. Its (Hopf) subalgebra generated by the twisted derivations, which can be interpreted as deformed translations, is dual to the associative algebra modeling the $\rho$-Minkowski space-time. The results are discussed in Section \ref{conclusion}.

\section{Twisted differential calculus}
\label{sec:section2}
\subsection{Star-product for $\rho$-Minkowski space-time}\label{section21}
\paragraph{}

We will use the star-product introduced in \cite{rho-1}. This latter, together with the associated involution is given by{\footnote{Our convention for the Fourier transform is $\mathcal{F}f(p) = \int \frac{d^dx}{(2\pi)^d}\ e^{- i p x} f(x)$ and $f(x) = \int d^dp\ e^{i p x} \mathcal{F}f(p)$. }}
\begin{align}
    (f\star g)(x_0,\vec{x},x_3)
    &=\int \frac{dp_0}{2\pi}\ dy_0\ e^{-i p_0 y_0} f(x_0 + y_0, \vec{x}, x_3) g(x_0, R(- \rho p_0) \vec{x}, x_3),
    \label{star-final} \\
    f^\dag(x_0, \vec{x}, x_3)
    &= \int \frac{dp_0}{2\pi}\ dy_0\ e^{- i p_0 y_0} \overline{f}(x_0 + y_0, R( - \rho p_0) \vec{x}, x_3),
    \label{invol-final}
\end{align}
for any functions $f,g$ in the associative $*$-algebra, denoted by $\mathcal{M}^4_\rho$ \cite{rho-1}, which describes the $\rho$-Minkowski space-time. In \eqref{star-final}, \eqref{invol-final}, $R(\rho p_0)$ denotes a $2 \times 2$ rotation matrix with dimensionless parameter $\rho p_0$, where $p_0$ and $\rho$ have respective mass dimension $1$ and $-1$.\\

The corresponding construction combines the main features of the group algebra related to the non trivial part of the coordinates algebra for the $\rho$-Minkowski space-time and the Weyl quantization map, as we will briefly recall and illustrate below. For more details, see e.g. \cite{physrep}. Note that a similar scheme, directly inherited from the old works of von Neumann and Weyl \cite{vonNeum, Weyl}, has already been applied to the $\kappa$-Minkowski space-time in \cite{DS}, \cite{PW2018}-\cite{MW2021}. The resulting star-product has been further used to build and study scalar field theories and gauge theories on this quantum space \cite{PW2018}-\cite{MW2021}, \cite{KMW-1}.\\

In the present situation, the relevant group is the special Euclidean group 
\begin{equation}
    \mathcal{G}_\rho
    := SE(2)
    = SO(2) \ltimes_\phi \mathbb{R}^2
    \label{g-rho}, 
\end{equation}
related to the non trivial part of the coordinate algebra for the $\rho$-Minkowski space-time given by 
\begin{equation}
  [x_0, x_1] = i \rho x_2,\ [x_0, x_2] = - i \rho x_1,\ [x_1, x_2] = 0,   
\end{equation}
which defines the Euclidean algebra $\mathfrak{e}(2)$. In \eqref{g-rho}, $\phi:SO(2)\to\text{Aut}(\mathbb{R}^2)$ denotes the action of any matrix of $SO(2)$ on elements of $\mathbb{R}^2$. \\
The convolution algebra is $\mathbb{C}(\mathcal{G_\rho}):=(L^1(\mathcal{G_\rho}),\circ,^\sstar)$ with convolution product $\circ$ together with involution $^\sstar$ given by
\begin{equation}
     (F\circ G)(s)
    = \int_{\mathcal{G}_\rho}\ d\mu(t) F(s t) G(t^{-1}),\ \ F^\sstar(x)= {\overline{F}}(x^{-1})\label{start-convolution}
\end{equation}
for any $F, G\in L^1(\mathcal{G}_\rho)${\footnote{Additionally, the functions must be compactly supported.}}, $s,t,x\in\mathcal{G}_\rho$, where ${\overline{F}}$ is the complex conjugate of $F$ and the Haar measure of $ \mathcal{G}_\rho $, $d\mu(t)$ reduces to the usual Lebesgue measure. \\

Next, assume that {\it{the elements of $\mathbb{C}(\mathcal{G}_\rho)$ are functions on a momentum space}}, i.e. write $F\in\mathbb{C}(\mathcal{G}_\rho)$ as $F=\mathcal{F}f$ where $\mathcal{F}$ is the Fourier transform. Then, upon parametrising any element of $\mathcal{G}_\rho$ as $( R(\rho p_0), \vec{p})$ where $\vec{p} \in \mathbb{R}^2$ and $R(\rho p_0)\in SO(2)$ denotes a $2 \times 2$ matrix with defining dimensionless parameter $\rho p_0$ with $p_0$ identified with the time-like component of a momentum $(p_0, \vec{p})$ and further using the defining group laws for $\mathcal{G}_\rho $ given by
\begin{align}
  (R(\rho{p_0}),\vec{p})\ (R(\rho q_0),\vec{q})
  &= \Big(R(\rho(p_0 + q_0)), \vec{p} + R(\rho{p_0})\vec{q} \Big), &
  \label{rho1} \\
  (R(\rho{p_0}), \vec{p})^{-1}
  &= \Big(R({-\rho p_0}), - R(-\rho{p_0})\vec{p} \Big), &
  \bbone = \bbone_2
  \label{rho2},
\end{align}
one can easily re-express \eqref{start-convolution} as
\begin{align}
    (\mathcal{F}f \circ \mathcal{F}g)(p_0, \vec{p})
    &= \int d^3q\ \mathcal{F}f \Big(R({\rho}({p_0+q_0})), \vec{p} + R({\rho}{p_0})\vec{q} \Big) \mathcal{F}g \Big(R({-{\rho}q_0}), - R({-{\rho}q_0}) \vec{q} \Big)
    \label{rho3} \\
    \mathcal{F}f^*(p_0,\vec{p})
    &= \overline{\mathcal{F}f} \Big(R({-{\rho}p_0}),-R({-{\rho}p_0})\vec{p} \Big),
    \label{rho4}
\end{align}
for any $\mathcal{F}f, \mathcal{F}g \in L^1(\mathcal{G}_\rho)$, where we used the fact that the Haar measure is the Lebesgue measure.\\

Now, introduce the Weyl quantization map $Q:\mathcal{M}_\rho^3\to\mathcal{B}({\mathcal{H}})$, a morphism of $\star$-algebra, where $\mathcal{M}_\rho^3$ is an associative $\star$-algebra of functions, with star-product $\star$ and involution $^\dag$. $Q$ is defined by 
\begin{equation}
    Q(f)
    = \pi(\mathcal{F}f)
    \label{weyl-operat}
\end{equation}
where $\pi:\mathbb{C}(\mathcal{G}_\rho)\to\mathcal{B}({\mathcal{H}})$ is known to be the induced $\star$-representation of $\mathbb{C}(\mathcal{G}_\rho)$ on $\mathcal{B}({\mathcal{H}})$ by some unitary representation of $\mathcal{G}_\rho${\footnote{One has $\pi(F)
   = \int_\mathcal{G} d\mu_\mathcal{G}(x) F(x) \pi_U(x)$ for any $F\in\mathbb{C}(\mathcal{G})$, where $\pi_U:\mathcal{G}\to\mathcal{B}({\mathcal{H}})$ is a unitary representation. Recall that $\pi$ is bounded and non degenerate.}}. Here, $\mathcal{B}({\mathcal{H}})$ denotes the algebra of bounded operators on some Hilbert space $\mathcal{H}$. As a morphism of $\star$-algebra, $Q$ satisfies
\begin{align}
    Q(f\star g)
    = Q(f) Q(g), &&
    (Q(f))^\ddag
    = Q(f^\dag),
    \label{defining-relations}
\end{align} 
where $(Q(f))^\ddag$ denotes the adjoint of $Q(f)$. In the same time, as induced $\star$-representation of $\mathbb{C}(\mathcal{G}_\rho)$, $\pi$ verifies\\
\begin{align}
    \pi(F\circ G)
    = \pi(F)\pi(G), &&
    \pi(F)^\ddag
    = \pi(F^\sstar)
    \label{pi-morph}.
\end{align}
The combination of \eqref{defining-relations} and \eqref{pi-morph} yields
\begin{align}
    f\star g
    = \mathcal{F}^{-1} (\mathcal{F}f \circ \mathcal{F}g), && 
    f^\dag
    = \mathcal{F}^{-1} (\mathcal{F}(f)^\sstar).
    \label{star-prodetinvol}
\end{align}
These relations, combined with \eqref{rho3}, \eqref{rho4} and finally adding a central element $x_3$, thus extending $\mathcal{M}_\rho^3$ introduced above to a new algebra $\mathcal{M}_\rho^4$ relevant to the 4-dimensional case, give rise after some computation to the star-product and involution \eqref{star-final}, \eqref{invol-final} constructed in \cite{rho-1}.\\

The above Weyl quantization framework selects out a natural trace which is defined here by the Lebesgue integral. Indeed, from
\begin{align}
    \int d^4x\ (f \star g^\dag)(x)
    = \int d^4x\ f(x) \overline{g}(x)
    \label{formule1}
\end{align}
which holds for any $f,g\in\mathcal{M}_\rho^4$, one infers that $\int d^4x\ (f\star f^\dag)(x)
    = \int d^4x\ f(x)\overline{f}(x)\ge0$ so that $\int d^4x$ defines a positive map $\int d^4x:\mathcal{M}_{\rho}^{4+}\to\mathbb{R}^+$ where $\mathcal{M}_{\rho}^{4+}$ is the set of positive elements of $\mathcal{M}_{\rho}^4$. \\
    Besides, one has
\begin{equation}
 \int d^4x\ (f\star g)(x)=\int d^4x\ (g\star f)(x),\label{cyclic}
\end{equation}
for any $f,g\in\mathcal{M}_\rho^4$, so that the trace is cyclic contrary to its counterpart in the case of $\kappa$-Minkowski space-time which is twisted \cite{PW2018}.\\
A convenient Hilbert product can be defined, namely
\begin{equation}
    \langle f,g\rangle :=\int d^4x\ (f^\dag\star g)(x)=\int d^4x\ \overline{f}(x)g(x),\label{zehilbertproduit}
\end{equation}
for any $f,g\in\mathcal{M}_\rho^4$, which coincides formally with the usual $L^2$ product. We will use this product to construct an gauge action functional in a while.\\

From now on, we denote generically by $\mathcal{M}_\rho:=(\mathcal{M}_\rho^4,\star,\dag )$ the associative $\star$-algebra modeling the $\rho$-Minkowski space-time. Note that the corresponding "algebra of coordinates" can be easily obtained from \eqref{star-final} and takes the expected form
\begin{align}
    [x_0, x_1] = i \rho x_2, &&
    [x_0, x_2] = - i \rho x_1, &&
    [x_1, x_2] = [x_1, x_3]=[x_3, x_2]=0.\label{coord-alg-intro}
\end{align}
For our present purpose, it will be sufficient to define $\mathcal{M}_\rho$ as the algebra of Schwartz functions equipped with the star-product and involution respectively given by \eqref{star-final} and \eqref{invol-final}. \\

\subsection{Derivations and $\rho$-deformed translations}\label{subsec:diif-cal}

We will use a framework directly adapted from the derivation-based differential calculus, see e.g. in \cite{dbv-1}. For past studies on noncommutative gauge theories based on this type of differential calculus, see \cite{cawa}, \cite{de_Goursac_2012}. \\

A convenient differential calculus which leads to a suitable commutative limit $\rho\to 0$ can be obtained by starting from the following set of twisted derivations of $\mathcal{M}_\rho$
\begin{equation}
    \mathfrak{D}=\big\{P_\mu:\mathcal{M}_\rho\to\mathcal{M}_\rho,\ \mu=0,3,\pm,\ \{ P_0,P_3 \}_{\bbone}\oplus \{ P_+\}_{\mathcal{E}_+}\oplus \{ P_- \}_{\mathcal{E}_-}\big\}\label{twisted-deriv},
\end{equation}
where, anticipating the discussion of Subsection \ref{subsection42}, $P_0$, $P_3$, $P_\pm=P_1\pm iP_2$ generates the Hopf algebra of $\rho$-deformed translations denoted by $\mathcal{T}_\rho$, which is a Hopf subalgebra of the $\rho$-deformed Hopf Poincar\'e algebra, denoted by $\mathcal{P}_\rho$ and the subscripts in \eqref{twisted-deriv} refer to the corresponding twist affecting the Leibnitz rule, namely
\begin{eqnarray}
P_i(f\star g)&=&P_i(f)\star g+f\star P_i(g),\ i=0,3\nonumber\\
P_\pm(f\star g)&=&P_\pm(f)\star g+\mathcal{E}_\mp(f)\star P_\pm(g)\label{twist-leib}
\end{eqnarray}
for any $f,g\in\mathcal{M}_\rho$ with
\begin{equation}
 \mathcal{E}_\pm=e^{\pm i\rho P_0}   \label{lestwists}.
\end{equation}

Eqn. \eqref{twist-leib}, \eqref{lestwists} are easily obtained by assuming that the Hopf algebra $\mathcal{T}_\rho$ acts on $\mathcal{M}_\rho$ as 
\begin{equation}
  (P_\mu \triangleright f)(x) = -i \partial_\mu f(x),\ \mu=0,3,\pm\label{action-pmu}  
\end{equation}
for any $f\in\mathcal{M}_\rho$ and then combining \eqref{action-pmu} with \eqref{star-final}. In Eqn. \eqref{action-pmu}, the symbol $\triangleright$ denotes the map 
\begin{equation}
    \triangleright:\mathcal{T}_\rho\otimes\mathcal{M}_\rho\to\mathcal{M}_\rho\label{cestlemap1}
\end{equation}
defining the action of $\mathcal{T}_\rho$ on $\mathcal{M}_\rho$.\\

In order to introduce quantum analogs of space-time symmetries, it is natural to require that $M_\rho$ behaves as a left module over a Hopf algebra corresponding to a deformation of the Poincar\'e algebra. \\
For convenience, we recall that given a Hopf algebra $H$ with coproduct $\Delta$ and counit $\epsilon$, a left $H$-module algebra, says $\mathcal{A}$, is an algebra with action map 
\begin{equation}
   \varphi:H\otimes \mathcal{A}\to\mathcal{A}, 
\end{equation}
(as e.g. \eqref{cestlemap1}) satisfying 
\begin{equation}
   \varphi\circ(\text{id}_H \otimes m) = m \circ (\varphi \otimes \varphi) \circ (\text{id}_H \otimes \tau \otimes \text{id}_\mathcal{A}) \circ (\Delta \otimes \text{id}_\mathcal{A} \otimes \text{id}_\mathcal{A}),  
\end{equation}
and 
\begin{equation}
  \varphi \circ (\text{id}_H \otimes 1_\mathcal{A}) = 1_\mathcal{A} \circ \epsilon,   
\end{equation}
where $m:\mathcal{A}\otimes\mathcal{A}\to\mathcal{A}$ is the product on $\mathcal{A}$, $1_\mathcal{A}: \mathbb{C} \to \mathcal{A}$ is the unit of $\mathcal{A}$ and $\tau : H \otimes \mathcal{A} \to \mathcal{A} \otimes H$ is the flip map.\\
Now, consider only the translations $P_\mu$. The inclusion of the rotation and boost part will be done in Section \ref{section4}. By further requiring that $\mathcal{M}_\rho$ is a left module algebra over $\mathcal{T}_\rho$, a standard algebraic computation using the above conditions yields the coproduct equipping the Hopf algebra $\mathcal{T}_\rho$, defined by $\Delta:\mathcal{T}_\rho\to\mathcal{T}_\rho\otimes\mathcal{T}_\rho$:
\begin{eqnarray}
    \Delta(P_i)&=&P_i\otimes\bbone+\bbone\otimes P_i,\ i=0,3\label{delta1}\\
    \Delta(P_\pm)&=&P_\pm\otimes\bbone+\mathcal{E}_\mp\otimes P_\pm\label{delta2}\\
    \Delta(\mathcal{E})&=&\mathcal{E}\otimes\mathcal{E}\label{delta3}
    \label{lescopro},
\end{eqnarray}
where we set
\begin{equation}
    \mathcal{E} := \exp(i \rho P_0). 
\end{equation}
Eqns. \eqref{delta1}, \eqref{delta2} expresse the compatibility between the star product \eqref{star-final} of the algebra $\mathcal{M}_\rho$ and the coproduct $\Delta$ of the Hopf algebra $\mathcal{T}_\rho$. \\
For computational purpose, note that the compatibility condition to be verified can be conveniently written as $t\triangleright(f\star g)=m(\Delta(t)(\triangleright\otimes\triangleright)(f\otimes g))$ for any $t\in\mathcal{T}_\rho$, $f,g\in\mathcal{M}_\rho$, where $m:\mathcal{M}_\rho\otimes\mathcal{M}_\rho\to\mathcal{M}_\rho$ defines as usual the star product as $m(f\otimes g)=f \star g$.\\

Finally, supplementing the pair $(\mathcal{T}_\rho,\Delta)$ with a co-unit $\epsilon:\mathcal{T}_\rho\to\mathbb{C}$ and an antipode $S:\mathcal{T}_\rho\to\mathcal{T}_\rho$
given by
\begin{equation}
\epsilon(P_\mu) = 0,\ \mu=0,3,\pm,\ \quad \epsilon(\mathcal{E}) = 1, \label{co-unitP}
\end{equation}
\begin{equation}
           S(P_0)=-P_0,\ S(P_3)=-P_3,\ S(P_\pm) = - \mathcal{E}_\mp P_\pm,\  \quad S(\mathcal{E}) = \mathcal{E}^{-1} \label{antipodeP}
\end{equation}
turns $(\mathcal{T}_\rho,\Delta,\epsilon,S)$ into a Hopf algebra. Notice that eqn. \eqref{antipodeP} can be obtained from the defining relation of the antipode, namely $m\circ(S\otimes\text{id})\circ\Delta=m\circ(\text{id}\otimes S)\circ\Delta=\epsilon$.\\

Upon using $(t\triangleright f)^\dag=S(t)\triangleright f^\dag$, for any $t\in\mathcal{T}_\rho$, $f\in\mathcal{M}_\rho$, combined with \eqref{antipodeP}, one obtains $(P_i\triangleright f)^\dag=P_i\triangleright (f^\dag)$, $i=0,3$, so that $P_0$ and $P_3$ behave as real derivations while $P_\pm$ are not since one has
\begin{equation}
    (P_\pm\triangleright f)^\dag=-\mathcal{E}_\pm P_\mp\triangleright f^\dag\label{pasreelderiv},
\end{equation}
for any $f\in\mathcal{M}_\rho$ where \eqref{antipodeP} has been used, which signals that $P_\mp$ are not real derivations.\\
Finally, one can easily check that $\langle P_\mu\triangleright f,g\rangle=\langle f,P_\mu\triangleright g\rangle$, $\mu=0,1,2,3$, for any $f,p\in\mathcal{M}_\rho$ where the Hilbert product is given by \eqref{zehilbertproduit} thus insuring that the $P_\mu$, $\mu=0,1,2,3$, define self-adjoint operators. \\

The Hopf algebra of deformed translations defined above will become a Hopf subalgebra of the $\rho$-deformed Poincar\'e algebra defined in Section \ref{section4}.

\subsection{Twisted differential calculus}
To introduce a noncommutative differential calculus, one has to equip $\mathfrak{D}$ \eqref{twisted-deriv} with a suitable algebraic structure. To do that, we first observe that one has obviously
\begin{equation}
    [P_\mu,P_\nu]=0,\ \mu,\nu=0,3,\pm, \label{crochet-lie}.
\end{equation}
Besides, one can easily verify that 
\begin{equation}
   (z.P_\mu)(f)=z\star P_\mu(f)=P_\mu(f)\star z:=(P_\mu .z)(f),\ \mu=0,3,\pm,  \label{relat-module}
\end{equation}
for any $z\in\mathcal{Z}(\mathcal{M}_\rho)$, $f\in\mathcal{M}_\rho$, where $\mathcal{Z}(\mathcal{M}_\rho)$ is the center of $\mathcal{M}_\rho$.\\
In order to deal with various derivations carrying different twists, see \eqref{twist-leib}, it is convenient to introduce an extra degree, called the twist degree. It is defined by
\begin{equation}
    \tau(P_i)=0,\ i=0,3,\ \tau(P_\pm)=\pm1\label{twistdegree}.
\end{equation}
Accordingly, the linear structures in \eqref{twisted-deriv} will be defined from {\it{homogeneous}} linear combinations of elements of \eqref{twisted-deriv}, i.e. all the elements in the linear combination have the same twist degree. This actually defines a grading which will extend to the differential calculus. It follows that $\mathfrak{D}$ \eqref{twisted-deriv} inherits a structure of graded abelian Lie algebra and graded $\mathcal{Z}(\mathcal{{M}_\rho})$-bimodule owing to \eqref{crochet-lie}, \eqref{relat-module} so that one can write in obvious notations
\begin{equation}
    \mathfrak{D}=\mathfrak{D}_0 \oplus \mathfrak{D}_+\oplus\mathfrak{D}_-\label{grading},
\end{equation}
where $\mathfrak{D}_i$, $i=0,\pm$ can be read off from \eqref{twisted-deriv}.\\

The notion of noncommutative differential calculus is not unique. In the rest of this subsection, we will present two differential calculi{\footnote{Note that $\mathfrak{D}$ does not involve all the derivations of $\mathcal{M}_\rho$ so that the present differential calculi is of the restricted type in the sense of \cite{dbv-1}.}} incorporating the various twists related to the derivations. \\
From now one, we set $\Omega^0(\mathcal{M}_\rho):=\mathcal{M}_\rho$. Formally, one would also have to define $\Omega^n_{(p,q,r)(\mathcal{M}_\rho)}$ as the set of $n$-linear forms (forms of degree $n$) where the subscripts denote the respective integer twist degrees for $\{P_0,P_3\}$, $P_+$, $P_-$ such that $p+q+r=n$. Then, the product of forms would amount to define a map such that  
\begin{equation}
   \times:\Omega^n_{(p_1,q_1,r_1)(\mathcal{M}_\rho)}\times\Omega^m_{(p_2,q_2,r_2)(\mathcal{M}_\rho)}\to\Omega^{n+m}_{(p_1+p_2,q_1+q_2,r_1+r_2)(\mathcal{M}_\rho)}.  
\end{equation}
In the following analysis, we will not have to use explicitly the twist degree otherwise than to restrict the linear structures of $\mathfrak{D}$ to homogeneous structures w.r.t this degree. To simplify the notations, we will omit from now on the corresponding subscripts.\\

To define a first suitable differential calculus, let us assume that the exterior algebra of differential forms is defined from the spaces of $\mathcal{Z}(\mathcal{M}_\rho)$-multilinear {\it{antisymmetric}} maps $\omega:\mathfrak{D}^n\to \mathcal{M}_\rho$ denoted by $\Omega^n(\mathcal{M}_\rho)$ according to the above remark. Let us define
\begin{equation}
   \Omega^\bullet(\mathcal{M}_\rho) 
    = \bigoplus_{n=0}^4 \Omega^n(\mathcal{M}_\rho).
\end{equation}
For any $n$-form $\omega\in\Omega^n(\mathcal{M}_\rho)$, one has:
\begin{eqnarray}
\omega(P_1,P_2,...,P_n)&\in&\mathcal{M}_\rho,\label{formule1}\\
\omega(P_1,P_2,...,P_n.z)&=&\omega(P_1,P_2,...,P_n)\star z\label{formule2},
\end{eqnarray}
for any $z$ in $Z(\mathcal{M }_\rho)$ and any $P_1,\dots,P_n\in\mathfrak{D}$, where in both \eqref{formule1}, \eqref{formule2}, the symbols $P_i$ stand for some derivations in $\mathfrak{D}$.\\

The product $\times: \Omega^\bullet(\mathcal{M}_\rho) \to \Omega^\bullet(\mathcal{M}_\rho)$ and differential $d$ are formally defined for any $\omega\in\Omega^p(\mathcal{M}_\rho)$, $\eta\in\Omega^q(\mathcal{M}_\rho)$ by
\begin{equation}
\omega\times\eta\in\Omega^{p+q}(\mathcal{M}_\rho)
\end{equation}
with
\begin{equation}
\begin{aligned}
    (\omega \times \eta) & (P_1, \dots, P_{p+q}) \\
    &= \frac{1}{p!q!} \sum_{\sigma\in \kS_{p+q}} (-1)^{{\sign}(\sigma)}
    \omega(P_{\sigma(1)}, \dots, P_{\sigma(p)}) \star \eta(P_{\sigma(p+1)}, \dots, P_{\sigma(p+q)}),
    \label{eq:form_prod}
\end{aligned}
\end{equation}
\begin{align}
\begin{aligned}
    \dd \omega(P_1, \dots, P_{p+1}) 
    =& \sum_{j = 1}^{p+1} (-1)^{j+1} P_j \triangleright\big( \omega( P_1, \dots, \vee_j, \dots, P_{p+1}) \big),
    \label{eq:koszul}
\end{aligned}
\end{align}
where the symbols $P_i$, $i=1,2,\dots, p+q$ in \eqref{eq:form_prod}, \eqref{eq:koszul} denote generically some derivations $P_\mu,\ \mu=0,3,\pm\in\mathfrak{D}$, ${\sign}(\sigma)$ is the signature of the permutation $\sigma$, $\kS_{p+q}$ is the symmetric group of $p+q$ elements and the symbol $\vee_j$ denotes the omission of the element $j$. The differential verifies
\begin{equation}
    \dd^2 = 0
    \label{eq:d2_0}.
\end{equation}
 \\ 
The differential satisfies a twisted Leibnitz rule which can be conveniently put into the form
\begin{equation}
  \dd(\omega\times\eta)=d\omega\times\eta+(-1)^{\delta(\omega)} \omega\times_\mathcal{E}d\eta \label{leibnitz-form}
\end{equation}
for any $\omega,\eta\in\Omega^\bullet(\mathcal{M}_\rho)$, where the symbol $\delta(.)$ denotes the degree of form while the symbol $\times_\mathcal{E}$ indicates that a twist will act on the first factor depending on the actual derivation acting on the 2nd factor in \eqref{leibnitz-form} which appears in the evaluation of forms on a suitable set of derivations. Indeed, one obtains the following evaluation $(\mathcal{E}(\omega)\times \dd\eta)(P_1,\dots,P_{p+q+1})$ which by combining \eqref{eq:form_prod} and\eqref{eq:koszul} with \eqref{leibnitz-form} gives rise to a combination of terms of the form
\begin{eqnarray}
&\ &\mathcal{E}(\omega)(P_{\sigma(1)},\dots,P_{\sigma(p)})\star \dd\eta(P_{\sigma^\prime(p+1)},\dots,P_{\sigma^\prime(p+q+1)})\nonumber\\
&=&\sum_j\mathcal{E}(\omega)(P_{\sigma(1)},\dots,P_{\sigma(p)})\star((-1)^{j+1} P_{\sigma^\prime(j)}\eta(P_{\sigma^{\prime}(p+1)},\dots,\vee_j,\dots ,P_{\sigma^\prime(p+q+1)}))\nonumber\\
&=&\sum_j\mathcal{E}_{\sigma^\prime(j)}\triangleright(\omega(P_{\sigma(1)},\dots,P_{\sigma(p)}))
\star((-1)^{j+1} P_{\sigma^\prime(j)}\eta(P_{\sigma^{\prime}(p+1)},...,\vee_j,..,P_{\sigma^\prime(p+q+1)}))\nonumber\\
\end{eqnarray}
where $\mathcal{E}_{\sigma^\prime(j)}$ is the twist linked to the derivation $P_{\sigma^\prime(j)}\in\mathfrak{D}$.\\

Hence, the data $(\Omega^\bullet(\mathcal{M}_\rho), \times, d)$ where the product and differential defined by \eqref{eq:form_prod} and \eqref{eq:koszul} defines the graded differential algebra of forms. Notice that one has 
\begin{equation}
    \omega \times \eta \ne (-1)^{\delta(\omega)\, \delta(\eta)} \eta \times \omega, 
\end{equation}
simply stemming from the noncommutativity of the product $\star$ in \eqref{eq:form_prod} so that the above algebra is not graded commutative.\\

Alternatively, another suitable differential calculus can be defined by modifying the definition of the product of forms, still preserving the above formal definition of the differential.\\

This other product is such that we recover the usual Leibnitz rule, i.e. the twists are included into the product. Namely, one obtains
\begin{equation}
  d(\omega\wedge\eta)=d\omega\wedge\eta+(-1)^{\delta(\omega)} \omega\wedge d\eta \label{leibnitz-form}
\end{equation}
for any $\omega,\eta\in\Omega^\bullet(\mathcal{M}_\rho)$ , where $\wedge$ is defined by
\begin{equation}
\begin{aligned}
    (\omega \wedge \eta) & (P_1, \dots, P_{p+q}) \\
    &= \frac{1}{p!q!} \sum_{\sigma\in \kS_{p+q}} (-1)^{{\sign}(\sigma)}
    (\mathcal{E}_{\sigma(p +1)}.. \mathcal{E}_{\sigma(p+q)} \triangleright \omega)(P_{\sigma(1)},..,P_{\sigma(p+1)}) \star \eta(P_{\sigma(p+1)},...,X_{\sigma(p+q)}).
    \label{eq:form_prod_2}
\end{aligned}
\end{equation}
The corresponding triple $(\Omega^\bullet(\mathcal{M}_\rho),\wedge,\dd)$ still defines a graded differential algebra which is still not (graded) commutative.

\section{Twisted connection and curvature}\label{section3}
Starting from $\mathfrak{D}$ \eqref{grading} together with a right hermitian module over $\mathcal{M}_\rho$, denoted by $\mathbb{E}$, a twisted connection can be defined as a map $\nabla:\mathfrak{D}\times\mathbb{E}\to\mathbb{E}$ satisfying linearity conditions and Leibnitz rule inspirated from the celebrated Koszul connection. In view of the discussion in Subsection \ref{subsec:diif-cal}, it will be convenient to consider the different actions of the map $\nabla$
\begin{equation}
    \nabla:\mathfrak{D}_i\times\mathbb{E}\to\mathbb{E},\ i=0,\pm
\end{equation}
satisfying for any $m\in\mathbb{E}$, $f\in\mathcal{M}_\rho$
\begin{equation}
\nabla_{P_\mu+P^\prime_\mu}(m)=\nabla_{P_\mu}(m)+\nabla_{P^\prime_\mu}(m),\ \forall (P_\mu,P^\prime_\mu)\in\mathfrak{D}_i\times\mathfrak{D}_i, \ i=0,\pm\label{sigtaucon1}
\end{equation}
\begin{equation}
  \nabla_{z.P_\mu}(m)=\nabla_{P_\mu}(m)\star z,\ \forall P_\mu\in\mathfrak{D},\forall z\in Z(\mathcal{M }_\rho)\label{sigtaucon2},
\end{equation}
\begin{equation}
    \nabla_{P_\mu}(m\triangleleft f)=\nabla_{P_\mu}(m)\triangleleft f+{\beta}_{P_\mu}(m)\triangleleft P_\mu(f),\ \forall P_\mu\in\mathfrak{D}\label{sigtaucon3},
\end{equation}
where $m\triangleleft f$ denotes the action of the algebra on the module and the linearity condition \eqref{sigtaucon1} holds for linear combinations of derivations homogeneous in twist degree.\\

We assume from now on that $\mathbb{E}$ is one copy of $\mathcal{M}_\rho$, that is
\begin{equation}
\mathbb{E}\simeq\mathcal{M}_\rho.\label{unecopy}
\end{equation}
Besides, we assume that the action of $\mathcal{M}_\rho$ on $\mathbb{E}\simeq\mathcal{M}_\rho$ is given by
\begin{equation}
m\triangleleft f=m\star f.\label{zeaktion}
\end{equation}

The map ${\beta}_{P_\mu}:\mathbb{E}\to\mathbb{E}$ in the Leibnitz rule \eqref{sigtaucon3} can be easily characterized by computing the left- and right-
hand side of the identity $\nabla_{P_\mu}((m\star f)\star g)=\nabla_{P_\mu}(m\star (f\star g))$ which is found to be verified provided 
\begin{equation}
    \beta_{P_\mu}=\mathcal{E}_\mu,\ \ \text{with}\ \ \mathcal{E}_\mu=\bbone, \bbone, \mathcal{E}_\pm,\ \ \mu=0,3,\pm\label{beta-fixe},
\end{equation}
so that, \eqref{sigtaucon3} becomes now
\begin{equation}
    \nabla_{P_\mu}(m\triangleleft f)=\nabla_{P_\mu}(m)\triangleleft f+\mathcal{E}_\mu(m)\triangleleft P_\mu(f),\ \forall P_\mu\in\mathfrak{D}\label{sigtaucon4}.
\end{equation}

Then, upon setting 
\begin{equation}
   A_\mu=\nabla_{P_\mu}(\bbone),\ \nabla_\mu:=\nabla_{P_\mu}
\end{equation}
which defines the gauge potential, the relation \eqref{sigtaucon4} yields
\begin{equation}
   \nabla_{\mu}(f)=A_{\mu}\star f+ P_\mu(f)  \label{covar-deriv}
\end{equation}
for any $f\in\mathcal{M}_\rho$.\\

In order to deal with a noncommutative analog of an hermitian connection, the right-module $\mathbb{E}$ is further assumed to be equipped with an hermitian structure. Recall that it is defined as a map $h:\mathbb{E}\times\mathbb{E}\to\mathcal{M}_\rho$ satisfying $h(m_1,m_2)^\dag=h(m_2,m_1)$, together with $h(m_1\star f_1,m_2\star f_2)=f_1^\dag\star h(m_1,m_2)\star f_2$ and $h(1,1)=1$, for any $m_1,m_2\in\mathbb{E}$ and any $f_1,f_2\in\mathcal{M}_\rho$. In the following, we use the canonical hermitian structure
\begin{equation}
    h(m_1,m_2)=m_1^\dag\star m_2\label{hermit-struc}.
\end{equation}
for any $m_1,m_2\in\mathbb{E}\simeq\mathcal{M_\rho}$. Then, compatibility between the connection and the hermitian structure can be expressed as twisted hermiticity conditions for $A_\pm$ and $A_0,A_3$ respectively given by
\begin{equation}
(h(\mathcal{E}_+\triangleright\nabla_+(m_1),m_2)
    +h(\mathcal{E}_+\triangleright m_1,\nabla_+(m_2)))+ (+\to -)=P_+h(m_1,m_2)+P_-h(m_1,m_2)\label{hermiticity-cond1},
\end{equation}
\begin{equation}
    h(\nabla_i(m_1),m_2)
    +h(m_1,\nabla_i(m_2)))=P_ih(m_1,m_2),\ i=0,3\label{hermiticity-cond2}
\end{equation}
for any $m_1,m_2\in\mathbb{E}$, which are verified provided
\begin{equation}
    A_\pm^\dag=\mathcal{E}_\pm\triangleright A_\mp,\ A_i^\dag=A_i,\ i=0,3\label{hermit-amu}
\end{equation}
The curvature, defined as a map 
\begin{equation}
    \mathcal{F}(P_\mu,P_\nu):=\mathcal{F}_{\mu\nu}:\mathbb{E}\to\mathbb{E},\ \mu,\nu=0,3,\pm 
\end{equation}
has the following expression
\begin{equation}
    \mathcal{F}_{\mu \nu} : = \mathcal{E}_\nu \nabla_\mu \mathcal{E}_\nu^{-1} \nabla_\nu - \mathcal{E}_\mu \nabla_\nu \mathcal{E}_\mu^{-1} \nabla_\mu,\ \ \mu,\nu=0,3,\pm\label{morph-curv}
\end{equation}
with $\mathcal{E}_\mu$ still given by \eqref{beta-fixe} and satisfies
\begin{equation}
    \mathcal{F}_{\mu\nu}=-\mathcal{F}_{\nu\mu},\ \mu,\nu=0,3,\pm,
\end{equation}
together with
\begin{equation}
    \mathcal{F}_{\mu\nu}(m\star f)=\mathcal{F}_{\mu\nu}(m)\star f,\label{module-morph}
\end{equation}
for any $m\in\mathbb{E}$, $f\in\mathcal{M}_\rho$, indicating that $\mathcal{F}_{\mu\nu}$ \eqref{module-morph} defines a morphism of module.\\

Set now 
\begin{equation}
  \mathcal{F}_{\mu\nu}(\bbone)=F_{\mu\nu}  
\end{equation}
which can be viewed as the noncommutative analog of the field strength. The corresponding "components" can be expressed as
\begin{equation}
F_{\mu \nu} = P_\mu A_\nu - P_\nu A_\mu + (\mathcal{E}_\nu\triangleright A_\mu) \star A_\nu - (\mathcal{E}_\mu\triangleright A_\nu) \star A_\mu,\ \ \mu,\nu=0,3\pm\label{fmunu}.
\end{equation}
By further using the convenient product of forms \eqref{eq:form_prod_2}, the 2-form curvature $F\in\Omega^2(\mathcal{M}_\rho)$ corresponding to \eqref{fmunu} can be expressed as
\begin{equation}
  F = \dd A + A \wedge A   \label{decadix}
\end{equation}
which thus satisfies the Bianchi identity
    \begin{equation}
    \dd F = F \wedge A - A \wedge F\label{bianchi},
\end{equation}
where $A\in\Omega^1(\mathcal{M}_\rho)$ is the 1-form connection.\\

The unitary gauge transformations are defined as the set of automorphisms of the module $\mathbb{E}$, says $\phi\in\text{Aut}(\mathbb{E})$, preserving the hermitian structure \eqref{hermit-struc}, which is expressed as $h(\phi(m_1),\phi(m_2))=h(m_1,m_2)$. By combining the fact that $\phi(m\triangleleft f)=\phi(m)\triangleleft f$ together with \eqref{zeaktion}, \eqref{hermit-struc} and setting $\phi(\bbone):=g\in\mathbb{E}$ one easily find that the group of unitary gauge transformations is given by
\begin{equation}
    \mathcal{U}=\{g\in\mathbb{E}\simeq\mathcal{M}_\rho,\ \ g^\dag\star g= g\star g^\dag= 1  \}.\label{groupdejauge}
\end{equation}
The twisted gauge transformations for the connection is 
\begin{equation}
    \nabla^g_\mu(.)=(\mathcal{E}_\mu\triangleright g^\dag)\star\nabla_\mu(g\star .)\label{gauge-connex},
\end{equation}
for any $g\in\mathcal{U}$, for which one easily find the gauge transformations of the gauge potential given by
\begin{equation}
A^g_\mu=(\mathcal{E}_\mu\triangleright g^\dag)\star A_\mu\star g+(\mathcal{E}_\mu\triangleright g^\dag)\star P_\mu g\label{gaugetrans-amiou},
\end{equation}
where $\mathcal{E}_\mu$ is still given by \eqref{beta-fixe}.\\

The corresponding twisted gauge transformations for the noncommutative field strength $F_{\mu\nu}$ are given by (no summation over indices $\mu,\nu$ in the RHS)
\begin{equation}
    F^g_{\mu\nu}=(\mathcal{E}_\mu\mathcal{E}_\nu\triangleright g^\dag)\star F_{\mu\nu} \star g 
    \label{transfmunu},
\end{equation}
for any $g\in\mathcal{U}$ and $\mathcal{E}_\mu$ as in \eqref{beta-fixe}.\\

At this stage, some comments are in order.\\
One can easily verify that
\begin{equation}
(\mathcal{E}_\mu\triangleright g)^\dag=\mathcal{E}_\mu\triangleright g^\dag.\label{starautom}
\end{equation}
 Thus, \eqref{starautom} shows that the twists $\mathcal{E}_\mu$ define $*$-automorphims of $\mathbb{E}\simeq\mathcal{M}_\rho$. Hence, the present framework underlying the gauge theory model on the $\rho$-Minkowski space-time space-time to be presented in the next section actually depends on two twists: one is trivial (the identity) while the other one is $\mathcal{E}_+$, with obvious inverse $\mathcal{E}_-$, is a $*$-automorphism of the module. \\
 
 The occurrence of $*$-automorphisms as twists in the present framework may be understood by comparing the algebraic structure underlying the present gauge theory framework to the one for the $\kappa$-Minkowski space-time considered in \cite{MW2020a}, \cite{MW2021}. There, the (unique) non-trivial twist given{\footnote{in the four dimensional case.}} by $\mathcal{E}=e^{-3P_0/\kappa}$ defines a regular automorphism, i.e., it satisfies $(\mathcal{E}\triangleright f)^\dag=\mathcal{E}^{-1}\triangleright (f^\dag)$. This latter is rigidly linked to the (one-parameter) group of $*$-automorphisms of the $\kappa$-Minkowski space 
 \begin{equation}
    \{\widetilde{\sigma}_t:=e^{it\frac{3P_0}{\kappa}} \}_{t\in\mathbb{R}},  \label{tomitagroup}
 \end{equation}
 forming the modular group for the KMS weight defined by the twisted trace occurring naturally in the very structure of the group algebra related to the $\kappa$-deformation considered in \cite{DS}, \cite{PW2018}, namely the (right) Haar measure of the affine group $\mathbb{R}^+\ltimes \mathbb{R}^3$. The above mentioned twisted trace verifies $\varphi(f\star_\kappa g):=\int d^4x\ f\star_\kappa g=\int d^4x\ (\mathcal{E}\triangleright g)\star f $ in obvious notations.\\
 
 In the present analysis, the affine group is simply replaced by the special Euclidian group $SO(2)\ltimes\mathbb{R}^2$ whose group factor $SO(2)$ acting on $\mathbb{R}^d$ can be viewed, up to technicalities, as a compactified version of the group factor $\mathbb{R}^+$ of the affine group. Roughly speaking, one passes 
 from $\mathbb{R}^+\simeq\{e^x \}_{x\in\mathbb{R}} $ to $SO(2)\simeq\{e^{ix} \}_{x\in\mathbb{R}} $. This extra $i$ factor in the argument of the exponential will also show up in the star product modeling the $\rho$-deformation of the Minkowski space (observe the arguments of the functions in \eqref{star-final}) and by duality will affect the coproduct equipping the (translation part of the) $\rho$-Poincar\'e algebra leading to the extra $i$ factor in the argument of the exponential defining the present twist $\mathcal{E}_+$, thus forcing $\mathcal{E}_+$ to be a $*$-automorphim. This twist is now an element of a group of $*$-automorphisms of $\mathcal{M}_\rho$ given by
 \begin{equation}
     \{\sigma_t:=e^{-it\rho P_0} \}_{t\in\mathbb{R}}.\label{decadix1}
 \end{equation}
  \\
 One might wonder if $\sigma_t$ \eqref{decadix1} plays a distinguished role in the present situation as $\widetilde{\sigma}_t$ \eqref{tomitagroup} does in the $\kappa$-Minkowski case \cite{MW2020a}, \cite{MW2021}. The fact that the present trace, $\omega(f):=\int d^4x\ f(x)$ for any $f\in\mathcal{M}_\rho${\footnote{For the present comment, it is again sufficient to restrict the functions of the algebra $\mathcal{M}_\rho$ to Schwartz functions. The extension to a suitable multiplier algebra can be carried out as it is done for the case of $\kappa$-Minkowski space.}}, has the standard property of cyclicity \eqref{cyclic} implies a negative answer, as it is obviously not a KMS weight. Indeed, recall that a KMS weight on a ($C^*$) algebra $\mathbb{A}$ for a modular group of $*$-automorphisms $\sigma_t$, $t\in\mathbb{R}$, is a positive linear map $\varphi:\mathbb{A}^+\to\mathbb{R}^+$ such that i) $\varphi\circ\sigma_z=\varphi$, ii) $\varphi(f^\dag\star f)=\varphi(\sigma_{\frac{i}{2}}(f)\star(\sigma_{\frac{i}{2}}(f))^\dag)$ where $\sigma_z$, $z\in\mathbb{C}$, is an analytic extension of $\sigma_t$ \cite{kuster}. The positivity of $\omega$ is obvious and the condition i) easily verified, since
$\omega\circ\sigma_z(f)=\int d^4x\ e^{-iz\rho P_0}\triangleright f=\int d^4x f(x)$. However, the condition ii) cannot be satisfied because the trace $\omega$ is now cyclic in view of \eqref{cyclic}. Simply compute $\omega(\sigma_{\frac{i}{2}}(f)\star(\sigma_{\frac{i}{2}}(f))^\dag)=\int d^4x\ e^{\frac{\rho}{2}P_0}(f)\star e^{-\frac{\rho}{2}P_0}(f^\dag)=\int d^4x\  f\star e^{-{\rho}{}P_0}\ne\int d^4x\ f^\dag\star f$.\\

Althought the present framework cannot accommodate a non-trivial KMS structure \cite{conn-rov}, the gauge theory obtained from the connection and curvature introduced in Section \ref{section3} has physically interesting properties. We turn now on to the corresponding construction.

\section{Gauge theory model on $\rho$-Minkowski space-time}\label{section4}
\subsection{Gauge invariant action}\label{subsec41}
The requirement that the formal commutative limit $\rho\to0$ of a 4-dimensional action describing a gauge theory on $\mathcal{M}_\rho$ coincides with the standard 4-dimensional QED action points toward the following candidate action
\begin{equation}
    S_\rho:=\frac{1}{4G^2}\langle F_{\mu\nu}, F_{\mu\nu}\rangle =\frac{1}{4G^2}\int d^4x\  F_{\mu\nu}^\dag\star F_{\mu\nu}=\frac{1}{4G^2}\int d^4x\ \overline{F_{\mu\nu}}(x) F_{\mu\nu}(x)\label{action-class},
\end{equation}
where $G$ is a dimensionless coupling constant, $F_{\mu\nu}$ given by \eqref{fmunu},  $\overline{f}$ denotes the complex conjugate of $f$ and summation over $\mu, \nu$ is of course understood; we used \eqref{zehilbertproduit} to obtain the third equality. It will be convenient to set $A_\pm:=A_1\pm iA_2$. \\
We will assume from now on that $A_1, A_2$ together with $A_0, A_3$ are real-valued.\\

It is easy to verify that the action $S_\rho$ is invariant under the gauge transformations \eqref{gaugetrans-amiou}, \eqref{transfmunu}. One uses the cyclicity of the trace together with
\begin{equation}
    ((\mathcal{E}_\mu\mathcal{E}_\nu)\triangleright g)^\dag=(\mathcal{E}_\mu
    \mathcal{E}_\nu)\triangleright g,\ \mu,\nu=0,3,\pm,
\end{equation}
for any $g\in\mathcal{U}$ \eqref{groupdejauge} and the fact that $\mathcal{E}_\mu\triangleright(f\star g)=(\mathcal{E}_\mu\triangleright f)\star(\mathcal{E}_\mu\triangleright g) $ combined with $g\star g^\dag=\bbone$ \eqref{transfmunu}.\\

At this stage, some comments are in order.\\

Note that the present $\rho$-Poincar\'e invariant (anticipating the result of Subsection \ref{subsection42}) gauge theory on $\mathcal{M}_\rho$ is 4-dimensional (and could be obviously constructed in principle in any $d\ge3$ dimensions). This has to be compared with the $\kappa$-Poincar\'e invariant gauge theory on $\kappa$-Minkowski \cite{MW2020a}, \cite{MW2021} for which the requirement of both gauge and $\kappa$-Poincar\'e invariance selects the dimension to take the unique value $d=5$, stemming from the fact that the natural trace involved in the action is twisted with a twist depending on the dimension. In the present situation, the trace is standard, having the usual cyclicity property so that no constraint on the dimension of the quantum space-time can occur.\\

From the rightmost equality in \eqref{action-class}, one easily realizes that the kinetic operator coincides with the one of (commutative) electrodynamics, as a mere consequence of \eqref{zehilbertproduit} and the structure of the differential calculus which is used. Namely, the quadratic kinetic term is 
\begin{equation}
   S_{\rho,kin}\sim\int d^4x\ A_\mu K_{\mu\nu}A_\nu,\ \ K_{\mu\nu}=\eta_{\mu\nu}\partial^2-\partial_\mu\partial_\nu,  
\end{equation}
where the summations run over $\mu,\nu=0,1,2,3$, the gauge-potential $A_\mu$ is real-valued (see beginning of Section \ref{subsec41}) in obvious notations. This is unlike the gauge model on $\kappa$-Minkowski derived in \cite{MW2020a}, \cite{MW2021} which involves a deformed kinetic operator, merely stemming from the difference between the twist structure characterizing each derivation based differential calculus.\\

\subsection{$\rho$-Poincar\'e invariance of the action}\label{subsection42}
We now examine the invariance of $S_\rho$ under the action of a Hopf algebra which defines a deformation of the usual (trivial Hopf structure of the) Poincar\'e algebra, denoted herafter by $\mathcal{P}_\rho$. We will show in particular that the associative algebra $\mathcal{M}_\rho$ modeling the $\rho$-Minkowski space-time is a left module algebra over $\mathcal{P}_\rho$. \\

In order to make contact with a deformation of the Poincar\'e algebra, we first supplement $\mathcal{T}_\rho$ with rotation and boost generators, i.e. we consider the enlarged set of generators $(P_\mu, M_j,N_j)$, $\mu \in \{0,+,-,3 \}$ and $j \in \{ +,-,3 \}$
where $M_j$ and $N_j$ denote respectively the rotations and boosts, assuming that these latter act on $\mathcal{M}_\rho$ as
\begin{eqnarray}
    (M_j \triangleright f)(x)& =& (\epsilon_{jk}^{l} x^{k}P_l \triangleright f )(x)\label{action-M} \\
    (N_j \triangleright f)& =& ( (x_0 P_j - x_j P_0) \triangleright f)(x)\label{action-N},    \end{eqnarray}
for any $f\in\mathcal{M}_\rho$, with $M_\pm=M_1+\pm iM_2$, $N_\pm=N_1\pm iN_2$, while \eqref{action-pmu} still holds where the map \eqref{cestlemap1} of course must now be extended to $\triangleright:\mathcal{P}_\rho\otimes\mathcal{M}_\rho\to\mathcal{M}_\rho$. From \eqref{action-pmu}, \eqref{action-M}, \eqref{action-N}, one infers that 
\begin{equation}
    \begin{split}
        &  [M_i,N_j] = i \epsilon_{ijk}N_k, \quad [M_i,M_j] = i \epsilon_{ijk} M_k, \quad [N_i,N_j] = - i \epsilon_{ijk} M_k, \quad [N_i,P_0 ] = iP_i \\
       &   [N_i, P_j] = i \delta_{ij}P_0, \quad [M_i,P_j] = i \epsilon_{ijk} P_k, \quad [P_\mu,P_\nu] = [M_j,P_0] = 0\label{rhom-Lie},
    \end{split}
\end{equation}
which defines the usual Lie algebra structure of the Poincar\'e algebra. In short, the Poincar\'e Lie algebra is not deformed.\\

We now show that the action \eqref{action-pmu}, \eqref{action-M}, \eqref{action-N} insures that $\mathcal{M}_\rho$ is a left module algebra over $\mathcal{P}_\rho$ equipped with a Hopf algebra structure.  \\
To do this, first compute the action of $M_j$ \eqref{action-M} on the star product \eqref{star-final}. Define the following left multiplication operator (i.e. usual product, not the star-product)
\begin{equation}
    L_{x_j}f:=x_jf,\ j=\pm,3,\ \ L_{x_0}f:=x_0f
\end{equation}
for any $f\in\mathcal{M}_\rho$ and set 
\begin{equation}
  x_\pm=\frac{1}{2}(x_1\mp ix_2), \ \ \partial_\pm=\partial_1\pm i\partial_2. \label{holomorph-var}  
\end{equation}
Then, a standard computation yields the following relations for any $f,g\in\mathcal{M}_\rho$:
\begin{eqnarray}
   L_{x_\pm}(f\star g)&=&(x_\pm f)\star g= (\mathcal{E}_\mp\triangleright f)\star (x_\pm g)\label{lem-1},\\
   L_{x_0}(f\star g)&=&(x_0f)\star g=f\star (x_0g)\label{lem-2},\\
   L_{x_3}(f\star g)&=&(x_3f)\star g=f\star (x_3g)\label{lem-3},\\
    L_{x_3}P_\pm\triangleright(f\star g)&=&(L_{x_3}P_\pm\triangleright f)\star g+
    (\mathcal{E}_\mp f)\star (L_{x_3}P_\pm\triangleright g)\label{lem-4},\\
    L_{x_\pm}P_3\triangleright(f\star g)&=&(L_{x_\pm}P_3\triangleright f)\star g+
    (\mathcal{E}_\mp f)\star (L_{x_\pm}P_3\triangleright g)\label{lem-5},\\
   L_{x_\pm}P_\mp\triangleright(f\star g)&=&(L_{x_\pm}P_\mp\triangleright f)\star g+
    f\star (L_{x_\pm}P_\mp\triangleright g)\label{lem-6} 
\end{eqnarray}
From \eqref{lem-1}-\eqref{lem-6}, one easily finds
\begin{eqnarray}
    M_\pm\triangleright (f\star g)  &=& (M_\pm\triangleright f)\star g+(\mathcal{E}_\mp\triangleright f)\star (M_\pm\triangleright g)\label{rotation-action},\\
    M_3\triangleright (f\star g)&=& (M_3\triangleright f)\star g+f\star (M_3\triangleright g)\label{m3-action},\\
    N_\pm\triangleright (f\star g)  &=& (N_\pm\triangleright f)\star g+(\mathcal{E}_\mp\triangleright f)\star (N_\pm\triangleright g)\label{rotation-action},\\
    N_3\triangleright (f\star g)&=&(N_3\triangleright f)\star g+f\star (N_3\triangleright g)\label{n3-action}.
\end{eqnarray}
Now, processing as in \ref{subsec:diif-cal}, the requirement that $\mathcal{M}_\rho$ is a left module algebra over $\mathcal{P}_\rho$ is verified if the coproduct $\Delta:\mathcal{M}_\rho\to\mathcal{M}_\rho\otimes\mathcal{M}_\rho$ is defined by
\begin{eqnarray}
\Delta(M_\pm)& = &M_\pm \otimes \bbone + \mathcal{E}_\mp \otimes M_\pm,\label{deltaM}\\
\Delta(N_\pm)& = &N_\pm \otimes \bbone + \mathcal{E}_\mp \otimes  N_\pm\label{deltaN},\\
\Delta(M_3)& = &M_3 \otimes \bbone + \bbone \otimes M_3\label{deltam3},\\
\Delta(N_3)& = & N_3 \otimes \bbone + \bbone \otimes N_3\label{deltan3},
\end{eqnarray}
while \eqref{delta1}-\eqref{delta3} still hold true for which however the map $\Delta$
extends to a coproduct map $\Delta:\mathcal{M}_\rho\to\mathcal{M}_\rho\otimes\mathcal{M}_\rho$. Finally, the counit and antipode are found to be
\begin{eqnarray}
    \epsilon(M_j)&=& \epsilon(N_j) = 0,\ j=\pm,3, \quad \epsilon(\mathcal{E}) = 1,\\ 
S(M_j) &= &- M_j, \quad S(N_j) = - N_j,\ j=\pm,3,
\end{eqnarray}
where again use has been made of the defining relation $m\circ(S\otimes\text{id})\circ\Delta=m\circ(\text{id}\otimes S)\circ\Delta=\eta \otimes\epsilon$, with unit $\eta: \mathbb{C} \to \mathcal{H} $, which must be supplemented by \eqref{antipodeP}, extended as a map $S:\mathcal{P}_\rho\to\mathcal{P}_\rho$.\\

From the above discussion, one concludes that $\mathcal{P}_\rho$ is a Hopf algebra involving $\mathcal{T}_\rho$ as a Hopf subalgebra, while $\mathcal{M}_\rho$ is a left module algebra over $\mathcal{P}_\rho$.  \\

In the present framework, it can be verified that $\mathcal{T}_\rho$ and $\mathcal{M}_\rho$ are dual as Hopf algebras. Note that viewing the algebra modeling the $\rho$-Minkowski space-time as the universal enveloping algebra of the Lie algebra of coordinates given by \eqref{coord-alg-intro} would imply that it supports a unique Hopf algebra structure defined by 
\begin{eqnarray}
    \Delta_{\mathcal{M}}(x_\mu)&=&x_\mu\otimes\bbone+\bbone\otimes x_\mu,\label{dual1}\\
   \epsilon_{\mathcal{M}}(x_\mu)&=&0,\ \ S_{\mathcal{M}}(x_\mu)=-x_\mu\label{dual2},
\end{eqnarray}
as a consequence of the universal property of the universal enveloping algebra of any Lie algebra.
Now, $\mathcal{T}_\rho$ and $\mathcal{M}_\rho$ are dual as bialgebras if 
\begin{eqnarray}
    \langle\Delta(t),x\otimes y\rangle&=&\langle t,x\star y \rangle=\langle 
    t_{(1)},x\rangle\langle t_{(2)},x \rangle\label{poule1}\\
    \langle ht,x\rangle&=&\langle h\otimes t,\Delta_{\mathcal{M}}(x)\rangle=\langle h,x_{(1)}\rangle\langle t,x_{(2)} \label{poule2}\rangle
\end{eqnarray}
for any $h,t\in\mathcal{T}_\rho$, $x,y\in\mathcal{M}_\rho$, where the bilinear map $\langle.,.\rangle:\mathcal{T}_\rho\times\mathcal{M}_\rho\to\mathcal{M}_\rho$ is the dual pairing and we used the Sweedler notation in the rightmost equalities \eqref{poule1}, \eqref{poule2}. Combining \eqref{poule1} with \eqref{holomorph-var} together with \eqref{twist-leib} and defining the dual pairing as
\begin{equation}
    \langle P_\mu,x_\mu\rangle=-i\delta_{\mu\nu}\label{zepairinghopf},
\end{equation}
a straightforward computation leads to the Lie algebra of coordinates \eqref{coord-alg-intro} (expressed in the $x_\pm,x_0,x_3$ variables), while similar consideration applied to \eqref{poule2} gives back to \eqref{dual1}. Finally, the bialgebra structures are extended to Hopf algebra structures by including the antipodes, which amounts to verify if
\begin{equation}
    \langle S(t),x\rangle=\langle t,S_{\mathcal{M}}(x)\rangle\label{ouf}
\end{equation}
holds true for any $t\in\mathcal{T}_\rho$, $x\in\mathcal{M}_\rho$ where $\langle.,.\rangle$ still denotes the above Hopf pairing \eqref{zepairinghopf}. The fact that \eqref{ouf} holds can be checked from a straightforward calculation with $S_\mathcal{M}$ given by \eqref{dual2}. Summarising the above discussion, one concludes that $\mathcal{T}_\rho$ and $\mathcal{M}_\rho$ are actually dual as Hopf algebras.\\

Going back to the gauge invariant action \eqref{action-class}, it turns out that $S_\rho$ is $\mathcal{P}_\rho$-invariant, which stems from the following relation
\begin{equation}
    h\blacktriangleright\int d^4x\ \mathcal{L}:=\int d^4x\ h\triangleright \mathcal{L}=\epsilon(h)\int d^4x\ \mathcal{L},\label{decadixII}
\end{equation}
which holds true for any $h\in\mathcal{P}_\rho$, $\mathcal{L}\in\mathcal{M}_\rho$. \\

To verify that \eqref{decadixII} is true, it is sufficient to compute the action of the generators of $\mathcal{P}_\rho$ on $\int d^4x\ \mathcal{L}$.\\
For $P_\mu$, $\mu=0,3,\pm$, one has 
obviously $\int d^4x\ P_\mu\triangleright\mathcal{L}=\int d^4x\ (-i\partial_\mu\mathcal{L})(x)=0$ for any $\mathcal{L}\in\mathcal{M}_\rho$, where we used \eqref{action-pmu}, so that the $P_\mu$'s verify \eqref{decadixII}. \\

In the same way, for $M_j$, $j=\pm,3$,  one has to compute various expressions of the form $I_{jk}=\int d^4x\ x_j\partial_k\mathcal{L}$ for $j\ne k$, see \eqref{action-M}. But $I_{jk}=\int d^4x\ \partial_k(x_j\mathcal{L})=0$ for any $\mathcal{L}\in\mathcal{M}_\rho$. A similar computation and conclusion holds for $N_j$, $j\pm, 3$. Hence, the generators of $\mathcal{P}_\rho$ satisfy \eqref{decadixII} which insures the $\rho$-Poincar\'e invariance of $S_\rho$.\\

\section{Discussion and conclusion}\label{conclusion}

In this paper, we have constructed a gauge theory on a particular deformation of the Minkowski space-time recently investigated in the literature \cite{marija1}, \cite{marija2}, \cite{fabiano2023bicrossproduct}, called the $\rho$-Minkowski space-time. As shown in \cite{rho-1}, it can be described by an associative algebra of functions $\mathcal{M}_\rho$ whose star-product is obtained from a combination of the Weyl quantization map and the defining properties of the convolution algebra of the special Euclidean group. This latter is simply the group related to the coordinate algebra for the $\rho$-Minkowski space-time. The algebra can be equipped with a natural trace which is simply defined by the Lebesgue integral, a feature which somewhat simplifies the implementation of the gauge invariance of the action.\\

The algebra for $\rho$-Minkowski $\mathcal{M}_\rho$ inherits a structure of left module over a Hopf algebra of twisted derivations which will form a Hopf subalgebra of a $\rho$-deformation of the Poincar\'e algebra. These twisted derivations are used to define suitable twisted noncommutative differential calculus underlying the gauge theory. \\
Then, the notion of twisted connection introduced in \cite{MW2020a}, \cite{MW2021} is adapted to the present situation involving derivations with different twists. We assume that the right hermitian module over $\mathcal{M}_\rho$ entering the definition of the connection is one copy of $\mathcal{M}_\rho$. The corresponding curvature is characterized and is found to satisfy a Bianchi identity. Recall that the presently used notion of connection can be viewed as a noncommutative extension of the Koszul description of a connection. The hermiticity condition obeyed by the (noncommutative) gauge potential is found to be twisted together with the gauge transformations.\\

Unlike the case of $\kappa$-Minkowski considered in \cite{MW2020a}, \cite{MW2021} within a similar approach based on Weyl map and group algebra (for the affine group), the twists appearing in the present analysis are now $*$-automorphims of the module instead of being regular automorphisms. As discussed at the end of Section \ref{section3}, this reflects the change in the structure of the group related to the coordinate algebra.\\

A four dimensional reasonable candidate for a noncommutative gauge theory on the $\rho$-Minkowski space-time can be easily obtained starting from the "square of the curvature", as given by eqn. \eqref{action-class} which has obviously a suitable commutative limit. The gauge invariance can be easily verified from \eqref{transfmunu} combined with the cyclicity of the trace represented by the Lebesgue integral. The kinetic operator of the classical action $S_\rho$ \eqref{action-class} coincides with the kinetic operator of standard electrodynamics. This stems from the Hilbert product \eqref{zehilbertproduit} combined with the twisted differential calculus used in this analysis. \\

As far as quantum symmetries of the $\rho$-Minkowski space-time $\mathcal{M}_\rho$ are concerned, it is found that $S_\rho$ is invariant under the action of a Hopf algebra, $\mathcal{P}_\rho$, which corresponds to a deformation of the Poincar\'e algebra which we have fully characterized in Subsection \ref{subsection42}. Note that within the present $\rho$-deformation, the Lie algebra structure of the Poincar\'e algebra remains undeformed. Besides, the Hopf subalgebra in $\mathcal{P}_\rho$ of the deformed translations can be verified to be dual of the $\rho$-Minkowski 
space-time.\\

Recently, other $\rho$-deformations of the Poincar\'e algebra have been considered from various viewpoints \cite{marija1}, \cite{marija2} and closely studied in \cite{fabiano2023bicrossproduct} where in particular two star products have been presented which however are different from the star product \eqref{star-final} used in this paper. It would be interesting to investigate the possible relationship between the Hopf algebra $\mathcal{P}_\rho$ and their counterparts considered in \cite{fabiano2023bicrossproduct}. Besides, further investigations on the quantum behaviour of the gauge theory considered in this paper are definitely needed in order to see if this gauge model suffers from the various pathologies, such as possible vacuum instability or UV/IR mixing, affecting most of the noncommutaitve gauge theories constructed so far. We will come back to these points in a forthcoming publication.

\section*{Acknowledgments}
JCW thanks P. Bieliavski and P. Martinetti for useful discussions at various stages of this work. He also thanks the Action 21109 CaLISTA ``Cartan geometry, Lie, Integrable Systems, quantum group Theories for Applications'', from the European Cooperation in Science and Technology.
\\

\printbibliography

\begin{comment}

\end{comment}

\end{document}